# CXR-Net: An Artificial Intelligence Pipeline for Quick Covid-19 Screening of Chest X-Rays

Haikal Abdulah[1,4‡], Benjamin Huber[1,4‡], Sinan Lal[1], Hassan Abdallah[3], Luigi L. Palese[4], Hamid Soltanian-Zadeh[5], Domenico L. Gatti[1,2,6*]

[1]Department of Biochemistry, Microbiology and Immunology, Wayne State Univ., Detroit, MI, USA
[2]NanoBioScience Institute, Wayne State Univ., Detroit, MI, USA
[3]Department of Biostatistics, University of Michigan, Ann Arbor, MI, USA
[4]Department of Basic Medical Sciences, Neurosciences and Sense Organs, Univ. of Bari Aldo Moro, Bari, Italy
[5]Departments of Radiology and Research Administration, Henry Ford Health System, Detroit, MI, USA
[6]Molecular Medicine Institute, Cambridge, MA 02138, USA

‡These authors, listed in alphabetical order, contributed equally to the study.

*E-mail: dgatti@med.wayne.edu

## Abstract

CXR-Net is a two-module Artificial Intelligence pipeline for the quick detection of SARS-CoV-2 from Antero/Posterior (A/P) chest X-rays (CXRs). Module 1 was trained on a public dataset of 6395 CXRs with radiologist annotated lung contours to generate masks of the lungs that overlap the heart and large vasa. Module 2 is a hybrid convnet in which the first convolutional layer with learned coefficients is replaced by a layer with fixed coefficients provided by the Wavelet Scattering Transform (WST). Module 2 takes as inputs the patients' CXRs and corresponding lung masks calculated by Module 1, and produces as outputs a class assignment (Covid *vs.* non-Covid) and high resolution heat maps that identify the SARS associated lung regions. Module 2 was trained on a dataset of CXRs from non-Covid and RT-PCR confirmed Covid patients acquired at the Henry Ford Health System (HFHS) Hospital in Detroit. All non-Covid CXRs were from pre-Covid era (2018-2019), and included images from both normal lungs and lungs affected by non-Covid pathologies. Training and test sets consisted of 2265 CXRs (1417 Covid –, 848 Covid +), and 1532 CXRs (945 Covid –, 587 Covid +), respectively. Six distinct cross-validation models with the same Covid +/– ratio, each trained on 1887 images and validated against 378 images, were combined into an ensemble model that was used to classify the CXR images of the test set with resulting Accuracy = 0.789, Precision = 0.739, Recall = 0.693, $F_1$ score = 0.715, $ROC_{AUC}$ = 0.852.

## 1. Introduction

A critical step in the fight against COVID-19 pandemic is the screening of infected patients and the rapid recognition of those affected by Severe Acute Respiratory Syndrome CoronaVirus 2 (SARS-CoV-2). The main screening method used for detecting COVID-19 infection is the reverse transcriptase-polymerase chain reaction (RT-PCR). The main screening methods for SARS-CoV-2 are chest X-ray (CXR) and computed tomography (CT). It was found in early studies that the majority of COVID-19 positive patients with pneumonia present bilateral abnormalities mostly in the form of ground-glass opacities and consolidations in CT and/or CXR images [1]. While CTs provide greater diagnostic accuracy, CXRs are more readily available and enable rapid triaging of patients in the most affected areas. Furthermore, since CXR imaging is typically carried out as part of a standard procedure for patients with a respiratory complaint, it represents an ideal complement to RT-PCR testing. Notwithstanding CXRs being among the most used imaging tests worldwide [2], with millions of CXRs generated annually, their correct interpretation is often a major challenge to radiologists, and researchers continue to develop methods of machine learning (ML) and Artificial Intelligence (AI) to assist in this task [3]. In particular, much effort is ongoing into the development of Neural Networks (NN) based AI diagnostic systems that can aid radiologists in accurately interpreting CXRs in SARS-CoV-2 cases [4, 5]. Here we present CXR-Net, a two-module pipeline for SARS-CoV-2 detection. Module I is based on Res-CR-Net, a type of NN originally developed for the semantic segmentation of microscopy images [6], which can process images of any size, retaining the original resolution of the input images in the feature maps of all layers and in the final output. Module I was trained on datasets of A/P CXRs with radiologist annotated lung contours to generate accurate masks of the lungs that either overlap or do not overlap the heart and large vasa, and are minimally influenced by regions of consolidation or other texture alterations from underlying pathologies. Module II is a hybrid convnet in which the first convolutional layer with learned coefficients is replaced by a layer with fixed coefficients provided by the Wavelet Scattering Transform (WST). A particular advantage of this hybrid net is the removal of network instabilities associated with adversarial images (noise or small deformations in the input images that are visually insignificant, but that the network does not reduce correctly, leading to incorrect classification). This net converges more rapidly than an end-to-end learned architecture, does not suffer from vanishing or exploding gradients, and prevents overfitting, leading to better generalization. Module II takes as inputs the patients' CXRs and corresponding lung masks calculated by Module I, and produces as outputs a class assignment (Covid *vs.* non-Covid) and high resolution heat maps that identify the SARS associated lung regions.

## 2. Methods

*2.1 CXR-Net Module 1: lung segmentation*

Lung segmentation is the process of semantic segmentation [7] that provides a mask of the lung regions, and by exclusion of the non-lung regions. Automated segmentation of the lungs in CXRs is a challenging problem due to the presence of strong edges at the rib cage and clavicle, the lack of a consistent lung shape among different individuals, and the presence of cardiovascular structures in the chest. When patients are healthy, because of the high contrast between the lung fields and their boundaries, ML/AI methods usually provide reliable segmentations. However, when patients have pathologies that produce lung density abnormalities, the lower contrast between the lungs and the surroundings makes the segmentation task in CXRs significantly more challenging [8].

Deep learning/Neural Network (DL/NN) approaches have become very popular to address these challenges, with the widespread use of encoder-decoder based convolutional neural networks (CNN) to perform lung segmentation (reviewed in [9, 10]). These method were typically tested with the Japanese Society of Radiological Technology (JSRT, [11, 12]), the Montgomery County (MC), and the Shenzen Hospital (SH) ([8,13], and https://www.kaggle.com/yoctoman/shcxr-lung-mask) datasets of segmented lungs. More recently, Carvalho Souza *et al.* [14] have proposed a method to incorporate pathology associated opacities as part of the lung segmentation by using a first NN to derive an initial segmentation that excludes the opacities, and a second NN to reconstruct the lung regions "lost" due to pulmonary abnormalities. Selvan *et al.* [15] have treated the high opacity regions as missing data and presented a modified CNN-based image segmentation network that utilizes a deep generative model for data imputation. Kholiavchenko *et al.* [16] have reported that further improvement in lung segmentation can be achieved by training the NN not only on the ground-truth segmentation masks, but also on the corresponding contours.

Recently, we have introduced Res-CR-Net [6] (https://github.com/dgattiwsu/Res-CR-Net), a new type of fully convolutional residual NN [17, 18], that departs from the widely adopted of U-Net models [19-22] with encoder-decoder architecture. Res-CR-Net combines residual blocks based on separable, atrous *c*onvolutions [23, 24] with residual blocks based on recurrent NNs [25]. This network displayed excellent performance when tasked to segment images from either electron or light microscopy in three/four separate categories, using only a small number of images for training [6]. CXR-Net Module 1 is a simplified version of Res-CR-Net which, despite lacking the recurrent NN blocks, achieves excellent performance in the lung segmentation of normal and pathologic CXRs.

*2.1.1 CXRs and lung segmentations sources*

The following CXR sources were merged to generate the databases used to train Module 1:

1. Japanese Society of Radiological Technology (JSRT) dataset [11]. The dataset consists of 247 posterior-anterior (PA) chest radiographs with and without chest lung nodules, with a resolution of 2048×2048, 0.175 mm pixel size, and 12-bit depth. The reference organ boundaries for the JSRT images for left and right lung fields, heart and left and right clavicles were introduced by van Ginneken *et al.* [12] in 1024×1024 resolution, and are available in the Segmentation in Chest Radiographs (SCR) database (https://www.isi.uu.nl/Research/Databases/SCR/).

2. Montgomery County (MC) dataset [13] (http://openi.nlm.nih.gov/imgs/collections/NLM-MontgomeryCXRSet.zip). This dataset, publicly available from the Department of Health and Human Services of Montgomery County (Maryland), was collected by the Montgomery County's Tuberculosis Control program and consists of 138 CXR images, of which 80 are from normal patients and 58 are from patients with some manifestation of tuberculosis. The CXR images are available in 12-bit gray-scale with resolutions of either 4020×4892 or 4892×4020 and 0.0875 mm pixel spacing in both horizontal and vertical directions. The MC dataset has lung segmentation masks excluding the heart and large vasa, which were marked under the supervision of a radiologist and made available by Candemir *et al.* [8].

3. Shenzhen Hospital (SH) dataset [13]. X-ray images in this dataset have been collected by Shenzhen No. 3 Hospital in Shenzhen, Guangdong providence, China (http://openi.nlm.nih.gov/imgs/collections/ChinaSet_AllFiles.zip). The X-rays were acquired as part of the routine care at Shenzhen Hospital. The set contains images in JPEG format. There are 326 normal X-rays and 336 abnormal X-rays showing various manifestations of tuberculosis. Lung segmentation masks, bounded by the heart and large vasa, were prepared manually by students and teachers of the Computer Engineering Department, Faculty of Informatics and Computer Engineering, National Technical University of Ukraine "Igor Sikorsky Kyiv Polytechnic Institute", Kyiv, Ukraine [26].

4. V7-Darwin dataset (https://darwin.v7labs.com/v7-labs/covid-19-chest-x-ray-dataset). This dataset contains 6500 images of AP/PA chest X-rays with pixel-level polygonal lung segmentations. 5863 images are sourced from https://data.mendeley.com/datasets/rscbjbr9sj/2 (also available and commonly referred to by the Kaggle dataset: https://www.kaggle.com/paultimothymooney/chest-xray-pneumonia/data). In addition to several images of normal lungs, the dataset includes 1970 images of viral pneumonia, 2816 images of bacterial pneumonia, 17 images of *Pneumocystis* pneumonia, 23 images of fungal pneumonia, 2 images of *Chlamydophila* pneumonia, and 11 images of unidentified pneumonia. Additional 517 cases of COVID-19 pneumonia are sourced from a collaborative effort [27] (https://github.com/ieee8023/covid-chestxray-dataset). Lung segmentations in this dataset were performed by human annotators and include most of the heart, revealing lung opacities behind the heart that may be relevant for assessing the severity of viral pneumonia. The lower-most part of the lungs is defined by the extent of the diaphragm, where visible. If the back of the lungs is clearly visible through the diaphragm, it is also included up until the lower-most visible part of the lungs. Uniformly shaped lungs also decouple the shape and content within the left lung from the size of the heart. Image resolutions, sources, and orientations vary across the dataset, with the largest image being 5600×4700 and the smallest being 156×156.

### 2.1.2 Image pre-processing

*2.1.2.1 Image resizing and ground truth labels.* In order to be compatible with the image format used locally for archiving CXRs, and to facilitate further processing by our NN, all images from the listed databases and the corresponding lung masks were resized to 300×340 pixels. All CXRs were histogram equalized to minimize differences in contrast/brightness among the datasets and within datasets. All ground truth masks of the lung regions in these CXRs were obtained from the sources listed above. When only a semantic mask of the lung region was available, a complementary mask of the non-lung regions was generated. At the end of this pre-processing step, all images and ground truth binary masks were of dimensions (300×340×1), with 2 masks (one for each class, lung and non-lung) per image.

*2.1.2.2 Training and validation datasets.* CXRs and corresponding lung segmentations sourced from the JSRT, MC, SH, and V7-Darwin datasets were further combined into two distinct training and validation sets:

1. JMS dataset. This dataset consists of 952 CXRs derived by combining the JSTR, MC, and SH datasets. Lung masks in this dataset exclude the heart and large vasa contours. The dataset was split into a training and a validation set, with 904 and 48 image/mask pairs, respectively.

2. V7 dataset. This dataset was derived from the original V7-Darwin dataset by removing all sagittal views and CT scans. It consists of 6395 CXRs, whose corresponding lung masks include the heart and large vasa contours. The dataset was split into a training and a validation set, with 6191 and 204 image/mask pairs, respectively.

In all cases in which patient identification was available, and multiple CXRs from the same patient were included in the database, the selection of samples for the training and validation sets was carried out by carefully avoiding any *leakage* (inclusion of CXRs from the same patient in both the training and validation set).

*2.1.2.3 Data augmentation.* To avoid overfitting, we relied on geometric data augmentation. Each pair of image and ground truth mask(s) was sheered or rotated at random angles, shifted with a random center, vertically or horizontally mirrored, and randomly scaled in/out. The parts of the image left vacant after the transformation were filled in with reflecting padding. During training, images in a batch were not shuffled, but each image underwent a different type of augmentation as determined by the consecutive calls of a random number generator starting from a fixed initial seed. The same type of augmentation was applied to an image and its segmentation masks.

*2.1.2.4 Module 1 Architecture*

A flowchart of Module 1 architecture is shown in **Fig. 1**.

A *Leaky ReLU* activation is used throughout. We have not noticed any improvement in segmentation accuracy by using the *ELU* activation, as reported by Novikoff *et al.* [28]. After the last residual block a *Softmax* activation layer is used to project the feature map into the desired segmentation.

The *Dice coefficient, D,* [29-33], and the *Dice loss, $L_D$*, defined as:

$$L_D(\hat{y}, y) = 1 - D = 1 - \frac{2\sum_i \hat{y}_i y_i + s}{\sum_i \hat{y}_i + \sum_i y_i + s}$$

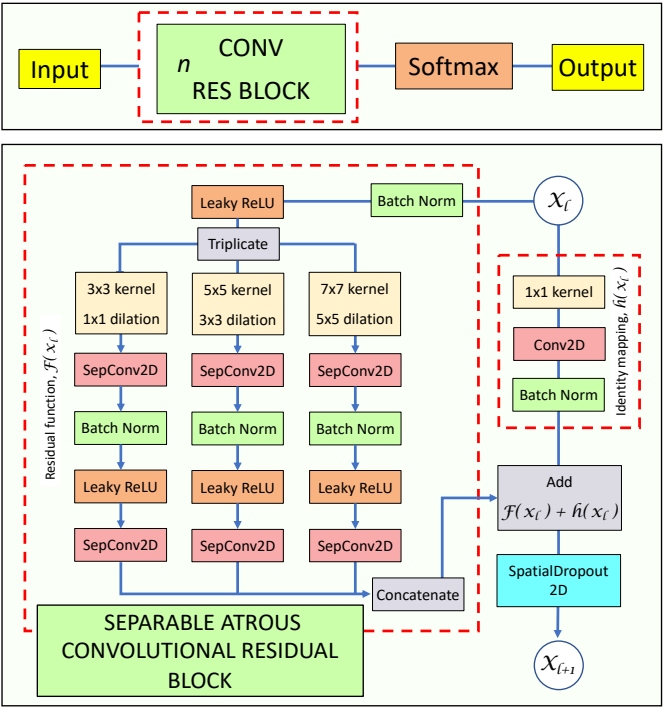

**Fig. 1.** Module 1 architecture. $n$ residual CONV RES blocks are repeated in a linear path along which the dimensions of the intermediate feature maps remain identical to those of the input image and of the output mask(s). The residual path of a CONV RES block consists of three parallel branches of separable/atrous convolutions that produce feature maps with the same spatial dimensions as the original image. Parallel branches inside the residual block are concatenated before adding them to the shortcut connection. A Spatial Dropout layer follows each residual block. In this study we have used 5 CONV RES blocks, each with kernel sizes of [3,3], [5,5], [7,7], and dilation rates of [1,1], [3,3], [5,5], respectively, with 16 filters in each residual branch, and 48 filters in the shortcut branch.

with $\hat{\mathbf{y}} \equiv \{\hat{y}_i\}$, $\hat{y}_i \in [0,1]$ being the probabilities for the $i$-th pixel, $\mathbf{y} \equiv \{y_i\}$, $y_i \in \{0,1\}$ being the corresponding ground truth labels, and $s$ a smoothing scalar, are often used for semantic segmentation tasks [34]. Loss functions of the *Dice* class containing squares in the denominator typically behave better in pointing to the ground truth, and even faster training convergence can be obtained by complementing the loss with a dual form that measures the overlap area of the complement of the regions of interest. These additional gains are implemented in the *Tanimoto loss* [22], defined as:

$$L_{\tilde{T}}(\hat{\mathbf{y}}, \mathbf{y}) = 1 - \tilde{T}(\hat{\mathbf{y}}, \mathbf{y})$$

where $\tilde{T}(\hat{\mathbf{y}}, \mathbf{y})$ is the *Tanimoto coefficient with complement*:

$$\tilde{T}(\hat{\mathbf{y}}, \mathbf{y}) = \frac{T(\hat{\mathbf{y}}, \mathbf{y}) + T(1 - \hat{\mathbf{y}}, 1 - \mathbf{y})}{2}$$

with $T(\hat{\mathbf{y}}, \mathbf{y})$ defined as:

$$T(\hat{\mathbf{y}}, \mathbf{y}) = \frac{\sum_i \hat{y}_i y_i + s}{\sum_i (\hat{y}_i^2 + y_i^2) - \sum_i \hat{y}_i y_i + s}$$

In this study we have used a *weighted Tanimoto loss* function throughout for training. Weights were derived with a *contour aware* scheme, by replacing a step-shaped cutoff at the edges of the mask foreground with a raised border that separates touching objects of the same or different classes [20]. The *unweighted Dice coefficient* defined above was used as the metric to evaluate segmentation accuracy.

### 2.2 CXR-Net Module 2: Covid vs. non-Covid classification

#### 2.2.1 CXRs sources

Antero-posterior CXRs from non-Covid and RT-PCR confirmed Covid patients were acquired at the Henry Ford Health System (HFHS) Hospital in Detroit (the 'HFHS dataset'). All non-Covid CXRs were from the pre-Covid era (2018-2019), and included images from both normal lungs and lungs affected by non-Covid pneumonia or other lung pathologies. CXR-Net Module 2 was trained against 2265 CXRs (1417 Covid –, 848 Covid +), and tested against 1532 CXRs (945 Covid –, 587 Covid +). Since multiple CXRs from the same patients were included in the database, CXRs were split into training and test set by carefully avoiding inclusion of CXRs from the same patient in both the training and test set. The training set, was further split for 6-fold cross-validation into 6 distinct sets of 1887 training images and 378 validation images, keeping in each set the same ratio of Covid + to Covid – images of the entire set of 2265 images. During training, CXRs were further assigned weights accounting for class imbalance (Covid – *vs.* Covid +).

#### 2.2.2 Image pre-processing

*2.2.2.1 Image resizing and lung mask generation.* HFHS CXR images were resized to 300×340 pixels and histogram equalized to minimize differences in contrast/brightness within the datasets, prior to passing through CXR-Net Module 1 to calculate the corresponding masks of the lung regions. At the end of this pre-processing step, all images and masks were of dimensions (300×340×1). Both training (2265 CXRs) and test images (1532 CXRs) were standardized to the common mean and standard deviation of the training set, mask values were in the [0,1] floating point range representing the probability of an image region to be part of the lungs. Standardized CXRs, and the corresponding masks and training weights were packed as H5 format files for input to CXR-Net Module 2.

*2.2.2.2 Data augmentation.* Each pair of image and Module 1 derived mask was identically sheered or rotated at random angles, shifted with a random center, vertically or horizontally mirrored, and randomly scaled in/out. The parts of the image left vacant after the transformation were filled in with reflecting padding. During training, images in a batch were not shuffled, but each image underwent a different type of augmentation as determined by the consecutive calls of a random number generator starting from a fixed initial seed.

*2.2.3 Module 2 Architecture*

Module 2 is a hybrid convnet in which the first convolutional layer with learned coefficients is replaced by a layer with fixed coefficients provided by the Wavelet Scattering Transform (WST) [35, 36]. A scattering network belongs to the class of CNNs whose filters are fixed as wavelets [37]. Thus, an important distinction between the scattering transform and a deep learning framework is that the filters are defined a priori as opposed to being learned. The construction of the scattering network relies on few parameters and is stable to a large class of geometric transformations [38], making its output a suitable generic representation of an image. A wavelet scattering framework enables the derivation of low-variance features from image data for use in machine learning and deep learning applications. These features are insensitive to translations of the input on an user-defined invariance scale and are continuous with respect to deformations. In the 2D case, these features are also insensitive to rotations. The scattering framework uses predefined wavelet and scaling filters. Efficient algorithms for the 2D scattering transform have been developed by Andén, Lostanlen, and Oyallon and implemented as an NN layer in Keras/Tensorflow via the Kymatio software [39].

A wavelet is an integrable and localized function in the Fourier and space domain, with zero mean. A family of wavelets is obtained by dilating a complex mother wavelet $\psi$ as $\psi_{j,\theta}$ where $j$ is a dilation scale and $\theta$ a rotation. $J$ and $L$ are integers parameterizing the dilation scale and the angular range. A scattering transform ($J$ being its spatial scale) generates features in an iterative fashion. Given a grayscale image $x$, and a local averaging filter $\varphi_J$ with a spatial window of scale $2^J$, we obtain the *zeroth order* scattering coefficients of $x$ as:

$$S^0 x = A_J x = x * \varphi_J$$

(where $*$ indicates here convolution). This operation leads to a down-sampling of scale $2^J$. For example, in the case of a grayscale image of dimensions $N \times N$, $S^0 x$ is a feature map of resolution $N/2^J \times N/2^J$ with a single channel. The zeroth order scattering transform is invariant to translations smaller than $2^J$ (which for this reason is often referred to as the *invariance scale* of the transform), but also results in a loss of high frequencies, which are necessary to discriminate signals. However, the information is recovered when computing the coefficients in the next stage. Each stage consists of three operations:

1. Take the *wavelet transform* of the input data with each wavelet filter in the filter bank. A wavelet transform, $W$, is the convolution of a signal with a family of wavelets, with an appropriate down-sampling. The *first order wavelet transform* of $x$ is:

$$W^1 x = x * \psi_{j_1, \theta_1}$$

2. Take the modulus of each of the filtered outputs.
3. Average each of the moduli with the scaling filter $\varphi_J$.

Thus, the *first order* scattering coefficients can be calculated as:
$$S^1 x = A_J |W^1| x = |x * \psi_{j_1, \theta_1}| * \varphi_J$$

In the grayscale image example of above, $S^1 x$ is a feature map of resolution $N/2^J \times N/2^J$ with $JL$ channels. Also in this first order, the use of averaging generates invariance to translation up to $2^J$. To recover some of the high-frequencies lost due to the averaging applied on the first order coefficients, we apply a second wavelet transform $W^2$ (with the same filters as $W^1$) to each channel of the first-order scatterings, *before* the averaging step.

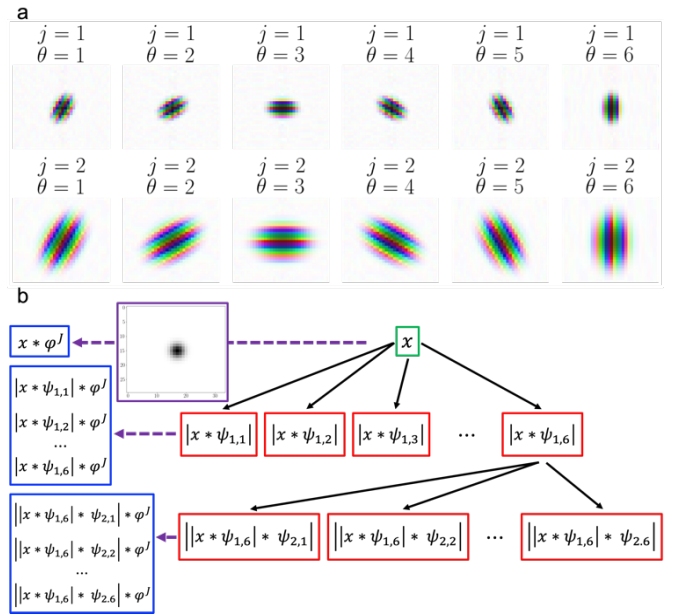

**Fig. 2.** Wavelet scattering transform. *Panel* a: Wavelets for each scales and angles used for the transform of an image $x$ of dimensions $30 \times 34$ with *J*=2 and *L*=6. Color saturation and color hue denote complex magnitude and complex phase, respectively. *Panel b*: a tree showing the calculation of the different channels of the scattering transform. Boxes with red outline represent the wavelet transform of the image with each wavelet filter and with modulus applied. The purple dashed arrows represent the averaging operation; the averaging filter is shown in the box with purple outline. The corresponding scattering coefficients for orders 0, 1, and 2 are shown in the boxes with blue outline. Additional scattering coefficients derived from the wavelet transform of the input image $x$ with the 2nd bank of wavelet filters are not shown.

This leads to the *second-order* scattering coefficients, calculated as:

$$S^2 x = A_J |W^2| |W^1| x = ||x * \psi_{j_1,\theta_1}| * \psi_{j_2,\theta_2}| * \varphi_J$$

In our grayscale image example, $S^2 x$ is a feature map of resolution $N/2^J \times N/2^J$ with $J(J-1)L^2$ channels. The energy of higher order scatterings rapidly converges to 0. Energy dissipation has a practical benefit, as it becomes possible to limit the number of wavelet filter banks in the framework with a minimal loss of signal energy. Thus, for most applications, a framework with two wavelet filter banks is sufficient. For this reason, the final scattering coefficient $S^J x$, which are the low-variance features derived from the image, correspond to the concatenation of the order 0, 1 and 2 scattering coefficients. $S^J x$ is a feature map with $1+JL+ 1/2 J(J-1)L^2$ channels, down-sampled by a factor of $2^J$ with respect to the original image. This representation has proven to linearize small deformations of images, to be non-expansive, and almost complete [40, 41], which makes it an ideal input to a deep convolutional network. An example, with $J=2$ and $L=6$, of the wavelets for each scales and angles that would be used for an image $x$ of dimensions $30 \times 34$ (1/10 of those used in this study) is shown in **Fig. 2a**. The corresponding *scattering transform* tree is shown in **Fig. 2b**.

A flowchart of the architecture of CXR-Net Module 2 is shown in **Fig. 3**. Module 2 takes as inputs the patients' CXRs and corresponding floating point lung masks (with [0,1] range representing the probability of an image region to be part of the lungs) calculated by Module 1, and produces as outputs a class assignment (Covid *vs*. non-Covid). High resolution heat maps that identify the SARS associated lung regions are calculated separately (see below). For input images of dimensions $300 \times 340$, the WAVELET SCATTERING TRANSFORM (WST) block, with $J=2$ and $L=6$, produces two outputs with down-sampling by $2^J = 4$. The 1st output is a feature map of dimensions $75 \times 85 \times 50$. The first 49 channels are the scattering transform. The 50th channel is the down-sampled floating point mask. The 2nd output is a binary mask of dimensions $75 \times 85 \times 1$ of the lung regions derived from the corresponding floating mask by thresholding at 0.5 and down-sampling by $2^J$. These outputs are passed to an ATTENTION block, containing two MultiHeadAttention layers [42, 43] that calculate a *cross-attention* feature map for the image rows using as *query* the first channel of the scattering transform, as *key* the binary mask, and as *value* the floating point mask. A *cross-attention* map for the image columns is obtained using the transposes of the same *query*, *key*, and *value* matrices. The two resulting attention maps are then multiplied pointwise by the *query* image producing a map representing a cross attention between the lung masks and the rows and columns of the down-sampled input CXR image (**Fig. 4**). This map is concatenated to the wavelet scattering transform output from the WST block producing a feature map of dimensions $75 \times 85 \times 51$. Upon

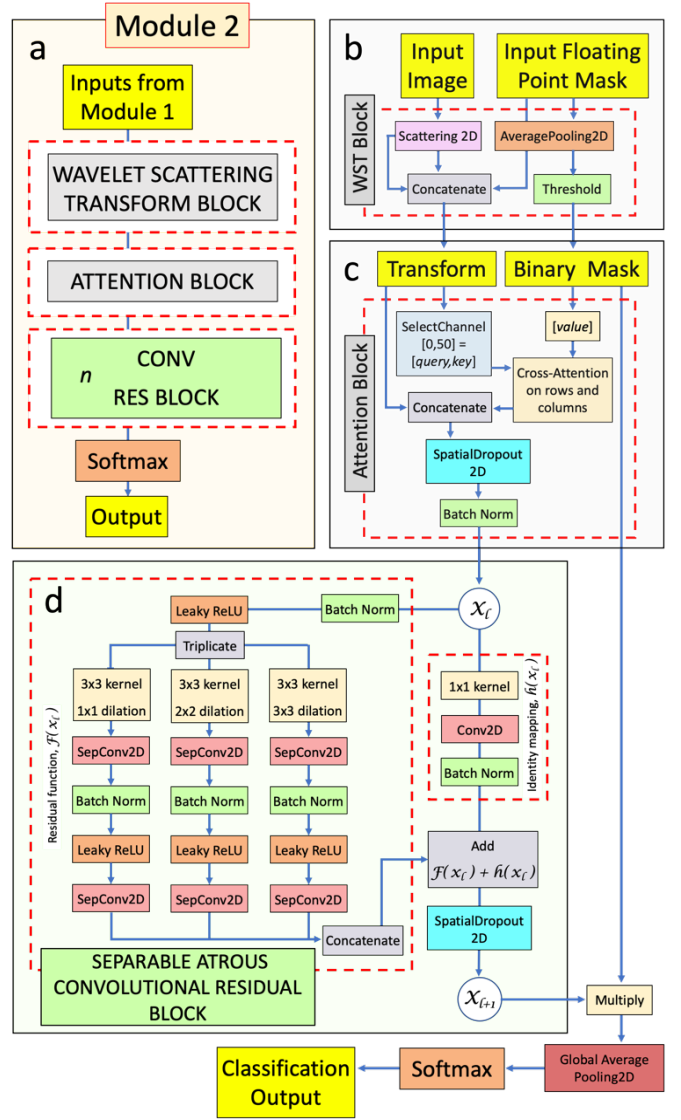

**Fig. 3.** Module 2 architecture. *Panel a*: overall architecture. *Panel b*: details of the WST BLOCK. *Panel d*: details of the ATTENTION BLOCK. The box labeled 'Cross-Attention on rows and columns' consists of two MultiHeadAttention layers working with the *query*, *key*, and *value* matrices for rows attention, and their transpose for columns attention. *Panel d*: details of the CONV RES blocks used in this module. In this case we have used 3 CONV RES blocks, each with kernel size of [3,3], and dilation rates of [1,1], [2,2], [3,3], respectively, with 17 filters in each residual branch, and 51 filters in the shortcut branch.

passing through a Spatial Dropout and a Batch Nomalization layer this map becomes the input to a residual CONV RES block (of the same architecture as described for Module 1) repeated 3 times in a linear path along which the dimensions of the intermediate feature maps remain identical to those of the input map. For these blocks, we have used a kernel size of [3,3] with dilation rates of [1,1], [2,2], [3,3], respectively, with 17 filters in each residual branch, and 51 filters in the shortcut branch. The final CONV RES block outputs a feature map of dimensions $75 \times 85 \times 2$, with channels corresponding to the

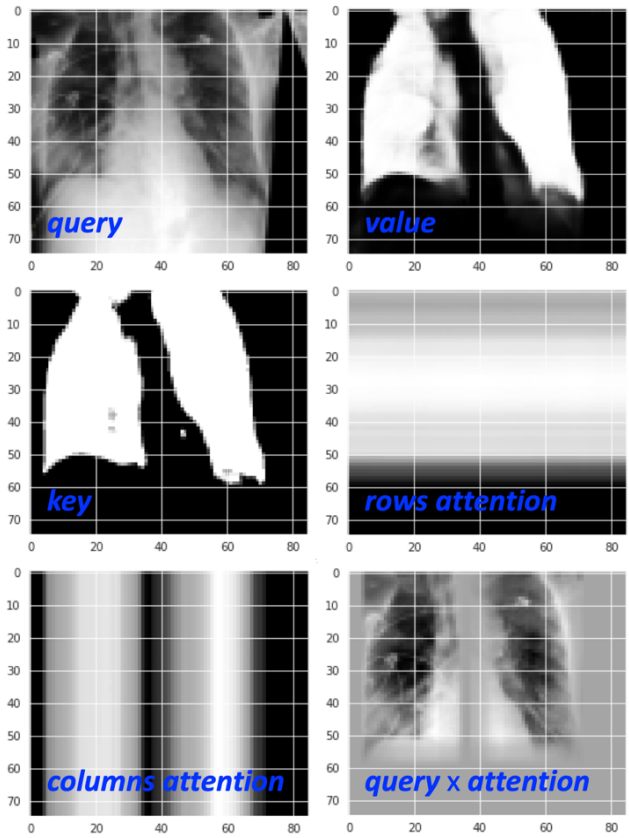

**Fig. 4.** *Cross attention*. The down-sampled CXR (top left), the lungs binary mask (middle left), and the lungs floating point mask (top right), are the *query*, *key*, and *value* features, respectively, used to calculate the cross-attention along the image rows (middle right) between the CXR and the lung regions in a Multi-head Attention layer (2 heads of size 64). Cross attention along the image columns (bottom left) is calculates using the *query*, *key*, and *value* transposes. Rows and columns cross attentions are multiplied pointwise by the *query* image to obtain the row- and column-wise CXR attention (bottom right).

Covid and non-Covid classes. Both channels are first multiplied by the binary mask of the lung region (2nd input to the CONV RES blocks), then globally averaged before passing to a Softmax layer for Covid vs. non-Covid classification.

In this study Module 2 was trained against the 2265 CXRs of the HFHS training set with 6-fold cross-validation. Individual models derived from each run were then combined into an *ensemble* model without averaging their layers coefficients. The final ensemble model (122,802 parameters) contained a single WST block followed by 6 parallel ATTENTION+CONV-RES blocks (**Fig. 5**).

Hybrid nets like CXR-Net Module 2 offer several appealing features. First, since the coefficient of the WST blocks are not learned, the net requires the refinement of fewer parameters. It has also been observed that the initial WST block helps removing deep network instabilities that may be associated with noise or deformations in the input images (including some produced by augmentation), which the network does not reduce correctly, and which may lead to incorrect classification [36]. In our experience, CXR-Net Module 2 converges more rapidly, suffer less from vanishing or exploding gradients, and leads to better generalization than an equivalent end-to-end learned architecture.

### 2.3 Software

CXR-Net was implemented using *Keras* [44, 45] deep learning library running on top of *TensorFlow* 2.2 [46], and is publicly available at https://github.com/dgattiwsu/CXR-Net. Training and testing were conducted at the High Performance Computing Grid of Wayne State University.

## 3. Results and Discussion

### 3.1 Module 1.

#### 3.1.1 Training with the JMS dataset

CXR-Net Module 1 was trained for 300 epochs with this dataset. Each epoch consisted of 113 batches of 8 images each. Thus, in every epoch the network trained on 904 different augmented images, and corresponding augmented masks as labels. Upon training Module 1 achieved ~97% segmentation accuracy on the images of the validation set. The training history showed no overfitting of the training set *vs.* the validation set (**Fig. 6a**).

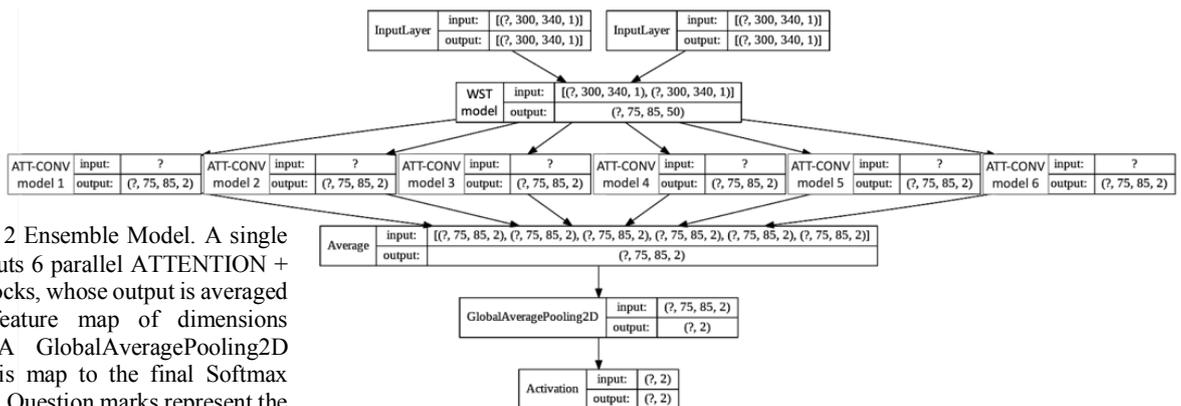

**Fig. 5.** Module 2 Ensemble Model. A single WST block inputs 6 parallel ATTENTION + CONV-RES blocks, whose output is averaged to a single feature map of dimensions 75 × 85 × 2. A GlobalAveragePooling2D layer passes this map to the final Softmax activation layer. Question marks represent the number of images in each batch.

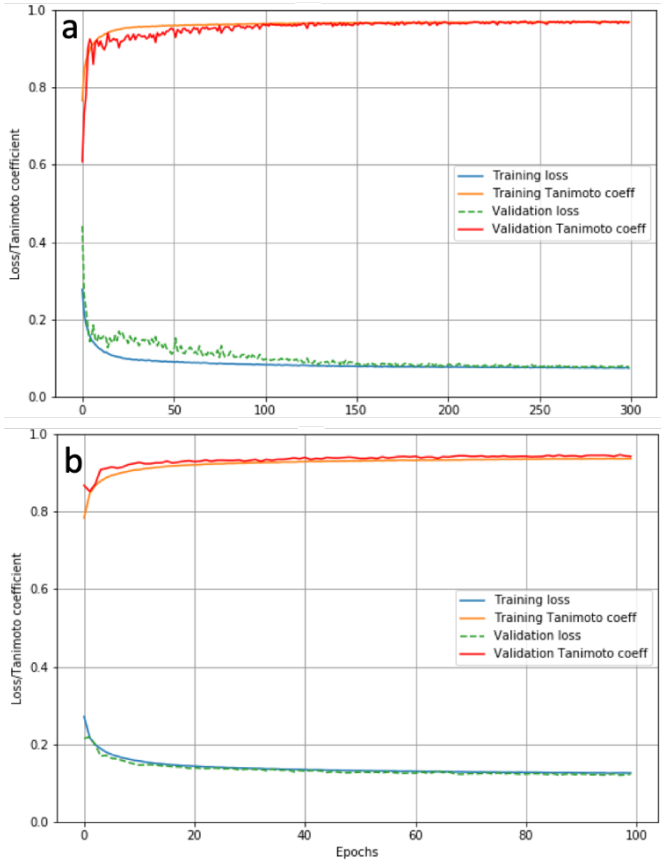

**Fig. 6.** Training and validation *weighted Tanimoto Loss* and *Tanimoto coefficient* vs. epochs for Module 1 processing of JMS (*Panel a*) and V7 (*Panel b*) images.

The segmentation task in these CXR images was to identify the regions occupied by the lungs with exclusion of other thoracic structures (skeletal and cardiovascular components) but including opacities due to the underlying pathology (**Fig. 7**). The total number of parameters refined was 59,165.

### 3.1.2 Training with the V7 dataset

Module 1 was trained for 100 epochs with this dataset. Each epoch consisted of 516 batches of 12 images each. Thus, in every epoch, the network trained on 6192 different augmented images, and the corresponding augmented masks as labels. Upon training, Module 1 achieved ~94% segmentation accuracy on the images of the validation set. As observed for the JMS dataset, the training history with the V7 dataset showed no overfitting of the training set *vs.* the validation set (**Fig. 6b**).

The segmentation task in these images was to identify the regions occupied by the lungs with exclusion of the skeletal structures visible in the CXR, but including cardiovascular components and opacities due to the underlying pathology (**Fig. 8**). Also in this case, the total number of parameters refined was 59,165.

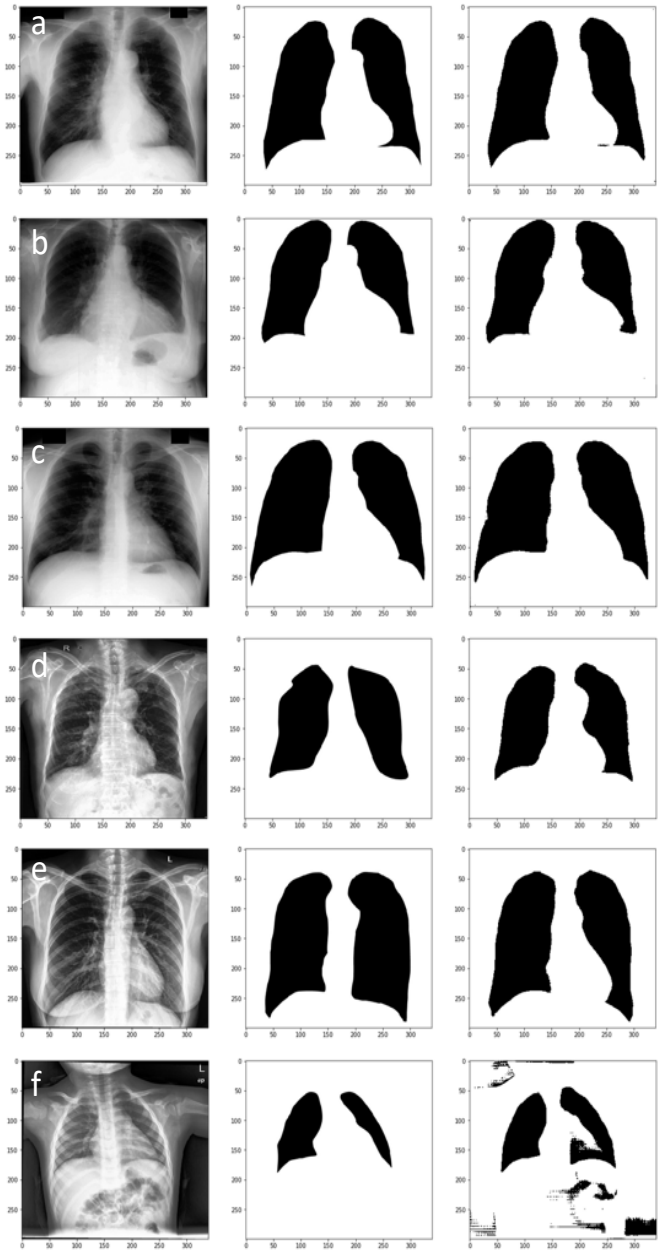

**Fig. 7.** Examples of lung segmentation in CXR images from the JMS validation subset. In each row of images: *Left panel*, CXR, *Center panel*, ground truth mask, *Right panel*, Module 1 predicted mask thresholded at 0.5 value. *Rows a-c* show three cases in which the mask is very similar to the ground truth mask. *Row d* shows an example in which the predicted mask is arguably more accurate than the ground truth mask. *Row e* shows an example in which the predicted mask excluded the heart from the lung segmentation, despite its incorrect inclusion by the annotator in the ground truth mask. *Row f* shows an example in which the predicted mask erroneously includes areas of the image that are either in the background or in the abdomen.

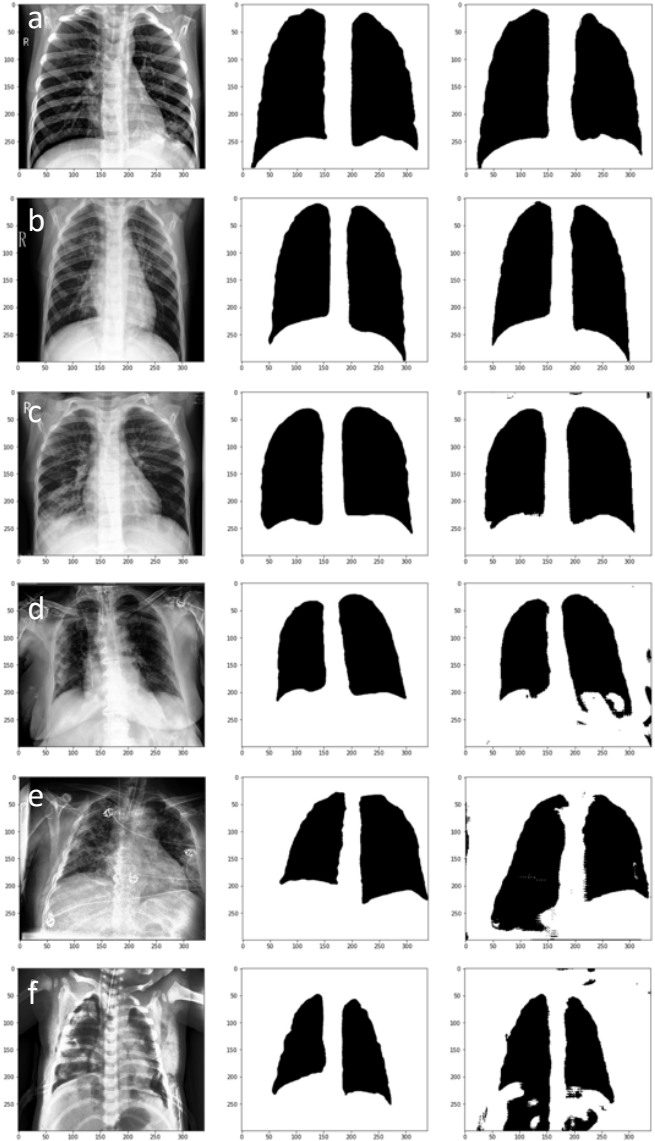

$$^aDice = \frac{2*TP}{(2*TP+FP+FN)}, ^bPrecision = \frac{TP}{(TP+FP)},$$
$$^cRecall = \frac{TP}{(TP+FN)}, ^dF_1\ score = \frac{2*precision*recall}{precision+recall}$$

with TP=*true positive*, FP=*false positive*, FN=*false negative*.

The performance of Module 1 with the JMS and V7 datasets is summarized in **Table 1** with respect to the metrics (*Dice coefficient, Precision, Recall, F₁ score*) most often used to evaluate neural networks performance.

### 3.1.3 Generation of lung masks for the HFHS dataset.

Automated segmentation of the lungs is usually considered a particularly difficult task when patients have pathologies that produce lung density abnormalities that decrease the contrast between lungs and non-lungs, or when such abnormalities are superimposed to the cardiovascular structures [8]. This becomes a particularly important issue when the final goal is the unsupervised recognition of the underlying pathology based on the density features of the lung tissue, as in the end these features tend to be excluded from the predicted masks leaving only the healthy regions of the lungs. Several different strategies have been reported to achieve this goal of correctly retaining abnormal tissue in the predicted lung masks. For example, Carvahlo Souza *et al.* [14] have used a separate NN to reconstruct the dense abnormalities that had been originally ignored, and had resulted in a considerable loss of segmented lung regions. Tang *et al.* [47] have used synthetic abnormal CXRs derived from adversarial training to train their segmentation model. More recently, Selvan *et al.* [15] have treated opacifications in CXRs as missing data to be inferred, and used a variational encoder to concatenate samples from the latent space to a standard CNN for segmentation.

In this study, Module 1 models trained on either the JMS (Module 1_JMS) or the V7 datasets (Module 1_V7) were used to generate lung masks for the CXRs in the HFHS dataset. Example of CXRs from this dataset, and the corresponding lung masks generated by Module 1_V7 are shown in **Fig. 9.** In all cases, mask values were kept in the original [0,1] floating point range representing the probability of an image region to be part of the lungs, without thresholding the mask at 0.5 value for conversion to a binary mask.

While CXR-Net Module 1 is a much simplified version of the Res-CR-Net [6]) originally designed for the semantic segmentation of microscopy images, it clearly exhibits very good performance also in the task of lung segmentation in CXRs from four different databases. In most cases, Module 1, trained on datasets that overlap (V7 dataset) or do not overlap the heart and large vasa (JMS dataset), was effective in achieving a semantic segmentation of the lungs that was almost indistinguishable from the ground truth masks produced by human annotation in the corresponding validation sets (**Figs. 7,8**). Its performance with the previously unseen CXRs of the HFHS dataset was also very good (**Fig. 9**), as judged by visual inspection of the masks, although in this case lung contours validated by radiologists were not available.

**Fig. 8.** Examples of lung segmentation in CXR images from the V7 validation subset. In each row of images: *Left panel*, CXR, *Center panel*, ground truth mask, *Right panel*, Module 1 predicted mask thresholded at 0.5 value. *Rows a-c* show three cases in which the mask is very similar to the ground truth mask. *Row d* shows an example in which the predicted mask is arguably more accurate than the ground truth mask. *Rows e-f* show two examples in which the predicted masks include areas of the images that belong to the abdomen.

**Table 1.** Performance metrics for CXR-Net Module 1 with the JMS and V7 validation datasets.

|  | $^aDice$ | $^bPrecision$ | $^cRecall$ | $^dF1$ |
| --- | --- | --- | --- | --- |
| JMS | 0.98 | 0.98 | 0.98 | 0.98 |
| V7 | 0.96 | 0.96 | 0.96 | 0.96 |

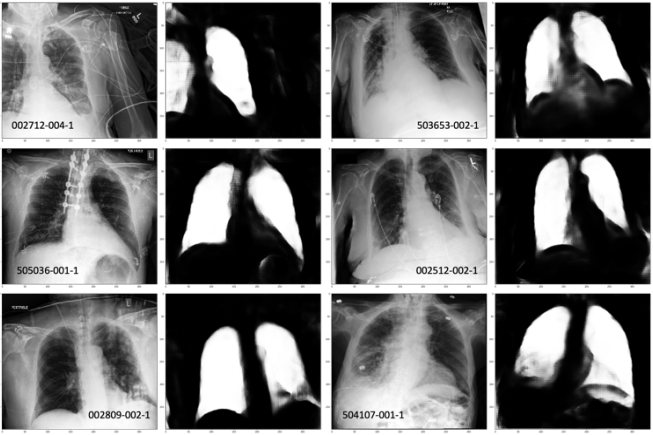

**Fig. 9.** Examples of lung segmentation in CXR images from the HFHS dataset. CXRs with HFHS numeric code starting with '00' and '50' were labeled by the radiologists as 'Covid +' and 'Covid –', respectively. Module 1_V7 generated floating point lung masks are shown next to each CXR.

While in this study we were only interested in lung segmentation, we anticipate that a comparable performance will be achieved also in multi-class segmentations (i.e., lungs, heart, vasa, claviculae). Furthermore, since CXR-Net Module 1 can easily process images of any size, without constrains imposed by the down-pooling and up-sampling operations of an encoder-decoder architecture, it is ideally suited to be included as a pre-trained module for *on the fly* segmentation of input CXRs in a classification network that seeks to identify lung pathologies.

*3.2 Module 2.*

*3.2.1 Training with the HFHS dataset*

The coefficients of the Wavelet Scattering Transform Block of Module 2 are fixed (non-trainable), and determined only by the scale and the number of rotations of the analytical wavelet (**Fig. 2**). An example of the output of this block with transform parameters $J=2$, $L=6$ (producing a feature map of dimensions $75 \times 85 \times 49$), with the concatenated floating point mask of the lungs ($75 \times 85 \times 1$) is shown in **Fig. 10**. The MultiHeadAttention layer of the following ATTENTION block, with 2 heads and *query*, *key*, and *value* vectors of 64 units, requires the training of ~1,100 parameters, and adds 1 extra channel to the resulting feature map. The CONV-RES blocks located further downstream require the training of ~20,000 parameters. All feature maps produced in each repeating block retain the same dimensions $75 \times 85 \times 51$) as the input map, with the exception of the last block, which produces a feature map of dimensions $75 \times 85 \times 2$ in preparation for the final GlobalAveragingPool2D and Softmax layers (**Fig. 3**).

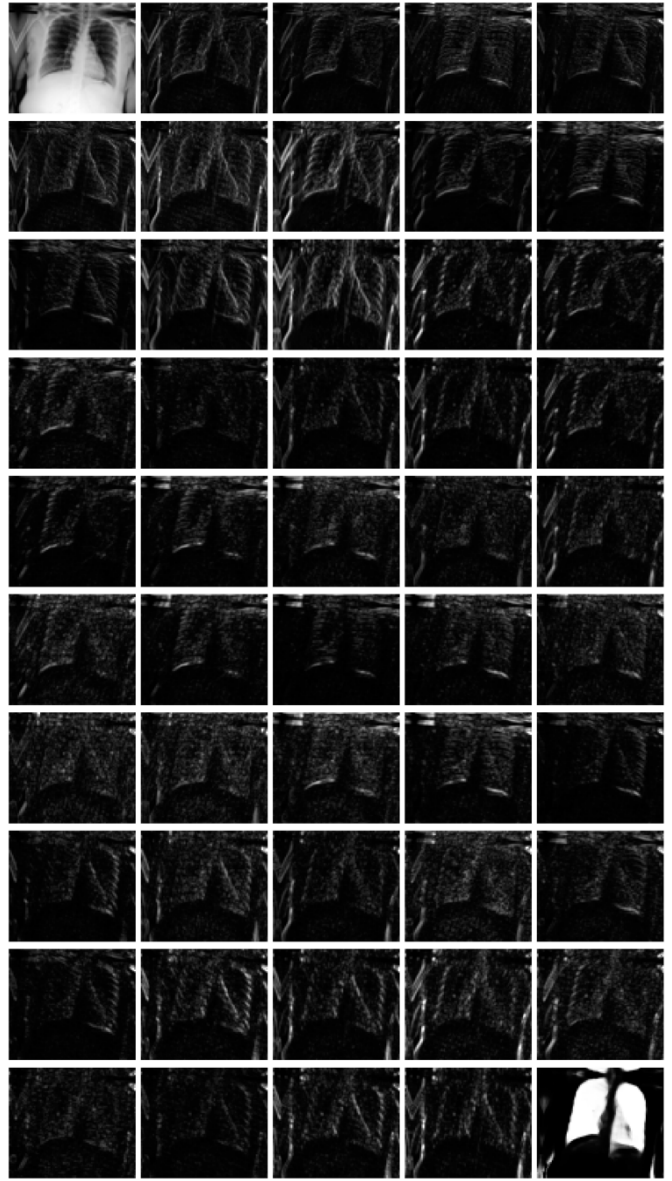

**Fig. 10.** Wavelet Scattering Transform. An example of the feature map of dimensions $75 \times 85 \times 49$ produced by the Wavelet Scattering Transform Block. The top-left panel is the input CXR, the bottom right panel is an extra channel with the floating point mask. All other panels represent the transform output at different scales and rotations of the analytical wavelet.

CXR-Net Module 2 was trained with 6-fold cross-validation with the HFHS training set of 2265 CXRs for 200 epochs in each validation run. Module 1_V7 lung masks were used. The random selection of the training and validation images in each run is shown in **Fig. 11a**. Each epoch processed 145 batches of 13 training images and 29 batches of 13 validation images; both training and validation images were augmented. The training and validation loss and accuracy for each run are shown in **Fig. 11b**. Receiver-Operating Characteristic (ROC) curves for the validation images in each run are shown in **Fig. 11c**; a confusion matrix combining all validation images is shown in **Fig. 11d**. Individual models of 20,467 parameters derived from each cross-validation run were combined into an *ensemble* model of 122,802 parameters without averaging their layers coefficients, as shown in **Fig. 5**. The Covid *vs.* non-Covid classification performance of the ensemble model with the non-augmented training set of 2265 CXRs and with the test sets of 1532 CXRs is displayed in terms of ROC curves and Confusion Matrices in **Figs. 11e,f,g,h**., and in terms of accuracy, precision, recall, and $F_1$ score in **Table 2**.

**Table 2.** Performance metrics of Module 2 model with the HFHS training, validation, and test sets.

|  | Mean (std) of 6 models | Ensemble model | |
|---|---|---|---|
|  | Validation sets | Training Set | Test Set |
| Accuracy | 0.813 (0.020) | 0.928 | 0.789 |
| Precision | 0.740 (0.056) | 0.878 | 0.739 |
| Recall | 0.793 (0.089) | 0.941 | 0.693 |
| $F_1$ score | 0.759 (0.025) | 0.908 | 0.715 |
| $ROC_{AUC}$ | 0.895 (0.007) | 0.981 | 0.852 |

*3.2.2 Heat maps of the lung regions*

We have used gradient weighted class activation heat maps (Grad-CAM saliency maps, [48]) to depict visually the decisions made by the ensemble model, so that the outcome of the model can be evaluated critically by a radiologist. The saliency maps highlight important areas that the model recognizes in the CXR, and uses to make its classification decision. It has been often found that Grad-CAM maps suffer from low resolution and cannot properly differentiate multifocal lesions within the image. This is likely due to the fact that most classification networks progressively shrink the spatial dimensions of the feature map, so that re-expansion of the gradient map from the final convolutional layer to the initial image dimensions inevitably leads to a smeared heat map. Ho *et al.* [49] have tried to overcome this problem by introducing a probabilistic Grad-CAM connected to a patch-based CNN. However, CXR-Net Module 2 does not suffer from a smearing effect because all intermediate feature maps produced at different layers of the module retain the same dimensions as the input image. We generated our saliency maps from the two channels of the final convolutional layer of Module 2, which are then globally averaged for direct input to the Softmax activation. The two channels contain the predictions for the CXR image probability of being Covid +

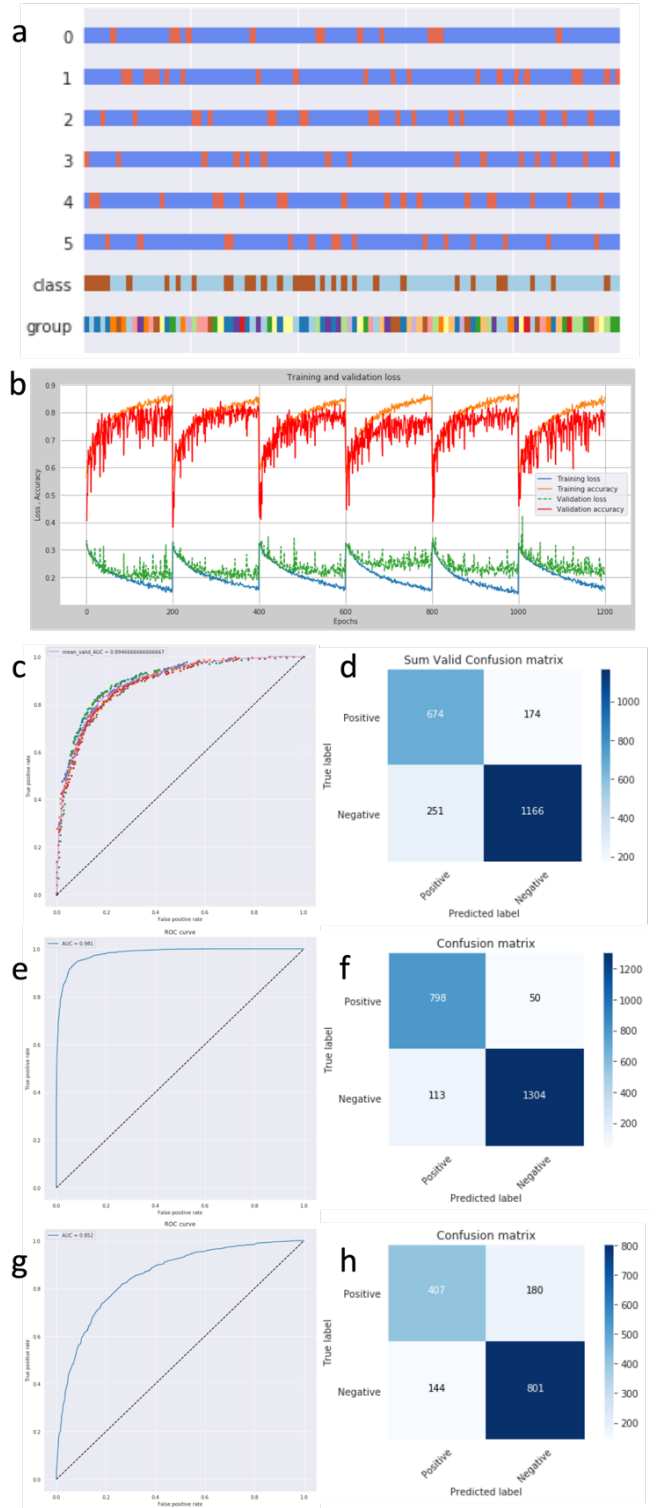

**Fig. 11.** Module 2 training with 6-fold validation. *Panel a*: partition of the training set of the HFHS dataset into 6 randomly selected train (blue sections) and validation (orange sections) subsets (rows labeled '0' to '5'). Cyan and brown sections (row labeled 'class') display the partition between Covid + and Covid – images. A multicolor row (labeled 'group') displays the partition of all images between different patients. *Panel b*: loss and categorical accuracy in the 6 training and validation sets. *Panel c*: validation ROC curves for each run (circles), and mean validation ROC curve for all 6 runs (continuous line). *Panel d*: confusion matrix for the combined validation sets. *Panels e,f*: ROC curve and confusion matrix for the Covid *vs* non-Covid classification by the ensemble model with the 2265 CXR images of the training set. In this case images were not augmented by translation, rotation, scaling, or mirror operations. *Panels g,h*: ROC curve and confusion matrix for the predictions by the ensemble model with the 1532 CXR images of the test set.

or Covid −, respectively, summing to 1. We generate Grad-CAM maps from both channels, and also a difference map between the two channels. When Module 2 clearly favors a Covid + or Covid − assignment, the difference map is almost identical to the map derived from the channel that produces the highest probability. When the assignment is uncertain (similar probabilities from both channels) the difference map tend to be featureless. When Module 2 classifies the image as Covid −, the difference maps identifies the lung regions that could still possibly be associated to a Covid form of pneumonia. Multifocal ground glass consolidations and other lung texture alterations were visualized effectively by our Grad-CAM maps. In particular, a noticeable Covid + activation map was observed only in Covid patients confirmed by RT-PCR, whereas almost no activations were observed in patients with no lung pathologies or with other non-Covid conditions (**Figs. 12,13**).

**Table 3.** Performance metrics of Module 2 model with the HFHS training, validation, and test sets when CXR regions below a given threshold value are not included in the Global Average Pooling of the final convolutional layer.

|  | Mean (std) of 6 models | Ensemble model | |
|---|---|---|---|
|  | Validation sets | Training Set | Test Set |
| Accuracy | 0.814 (0.022) | 0.938 | 0.793 |
| Precision | 0.728 (0.041) | 0.893 | 0.738 |
| Recall | 0.810 (0.075) | 0.952 | 0.709 |
| $F_1$ score | 0.764 (0.043) | 0.921 | 0.723 |
| $ROC_{AUC}$ | 0.897 (0.017) | 0.985 | 0.850 |

However, it is important to keep in mind that there might be significant differences in the rational applied by a radiologist and by an AI system in deciding which regions of a CXR are more important to classify an image as Covid + or Covid −. For example, if in SARS-CoV-2 some specific regions of the lungs are less likely to present texture alterations, while other forms of pneumonia may more uniformly affect the entire lung, an AI algorithm could produce saliency maps that assign similar scores to lung regions that *do not show* evidence of consolidation, and to those a radiologist would recognize as Covid related alterations of texture. We have explored the role of *clear* lung regions (as determined by an appropriate thresholding of the input CXRs) in the decision process of Module 2 by excluding these regions from the GlobalAveragePooling2D layer that feeds the final Softmax classification. This approach is conceptually similar to that of multiplying the gradient-based map by the input image [50-52], except that we multiply by a binary mask of the input image the feature map of the last convolutional layer rather than its gradient map. In this case, we have observed a larger spread in the performance of the 6 validation runs, with a marginal improvement of the mean performance metrics of the validation models, and in some categories (i.e., *accuracy, recall, $F_1$ score*) of the ensemble model (compare **Table 2** to **Table 3**). Accordingly, we have observed a small improvement in some of the saliency maps that were incorrectly assigned when not using thresholding (**Fig. 14**).

Since the amount of thresholding required to achieve this small improvement may depend on the data set, it is not the default mode of Module 2, but is available as a user defined option. Alternatively, Module 2 can multiply the final convolutional layer by the input CXR image, rescaled to pixel values in the interval [0,1]. This operation down-weighs the clear regions of the lungs, and up-weighs their denser counterparts in the decision making process of Module 2. Also in this case we have observed that some saliency maps show better correspondence with the radiologist assignment, although the overall classification performance is slightly degraded.

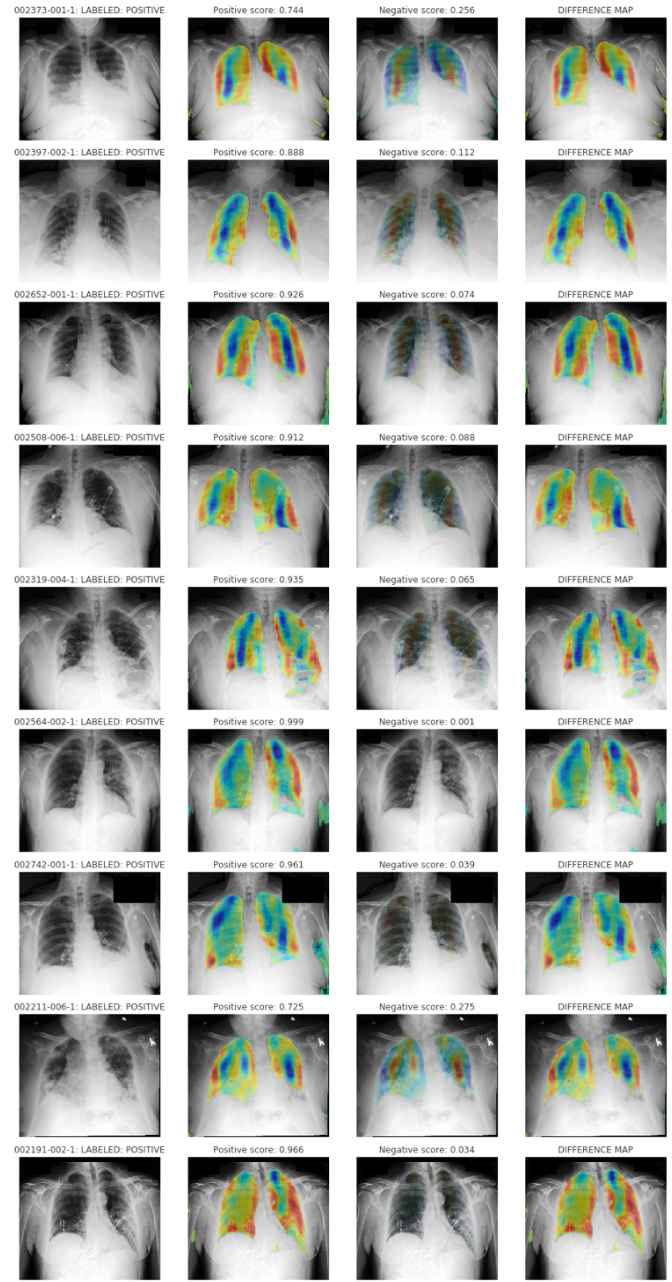

**Fig. 12.** Ensemble Module 2 predictions and Grad-CAM maps for Covid + CXRs in the HFHS test set. *Panels of each row, from left to right*: 1. input CXR and radiologist assignment based on a combination of the image analysis and RT-PCR positivity; 2. Positive score and corresponding Grad-CAM map (from channel 1 from the last convolutional layer); 3. Negative score and corresponding Grad-CAM map (from channel 2 from the last convolutional layer); 4. Difference map between the two Grad-CAM maps derived from the two channels of the last convolutional layer.

## 4. Conclusions

In this report we have presented CXR-Net, a neural network featuring a dual-module architecture for the sequential semantic segmentation of the lung fields in A/P CXRs, and their classification as either non-Covid (normal or other non-Covid related pathology) or SARS-CoV-2. In most cases CXR-Net Module 1 was effective in achieving a semantic segmentation of the lungs that was almost indistinguishable from the ground truth masks produced by human annotation (**Figs. 7,8**).

Module 1 architecture offers some advantages with respect to the traditional encoder-decoder architecture, as its layers contain no pooling or up-sampling operations, and therefore the spatial dimensions of the feature maps at each layer remain unchanged with respect to those of the input images and of the segmentation masks used as labels or predicted by the network. For this reason, Module 1 is completely modular, with residual blocks that can be proliferated in a straight down linear fashion as needed (**Fig. 1**), and it can process images of any size and shape without changing layers size and operations.

CXR-Net Module 2 is a novel type of convolutional classification network for biomedical images, in which the initial convolutional layers with learned parameters are replaced by the Wavelet Scattering Transform (WST) of the input image, with fixed parameters determined only by the scales and rotations of the analytical wavelet used in the transform [35-37]. The use of the WST as an image pre-processing step is equivalent to a process of transfer learning in which the input image is first passed through a CNN pretrained on a very large number of unrelated images, in order to provide an initial feature map that generalizes well also to the specific case at hand. The WST block of CXR-Net Module 2 produces a feature map that is passed on to a CNN with the same general architecture as Module 1, so that all intermediate feature maps maintain the same dimensions as the initial input image. In this study, we show how this type of hybrid network achieves excellent classification performance with only a small number of parameters. CXR-Net module 2 requires the refinement of only ~21,000 parameters for images of dimensions $300 \times 340$. For this reason, the models derived from several independent cross-validated training runs of Module 2 can be combined into a single *ensemble model* without parameter averaging, that still uses an overall small number of parameters. In this study, we have combined 6 cross-validated runs of Module 2 into an ensemble model with only ~123,000 parameters. In our experience, retaining the individual Module 2 validation models as parallel paths, rather than averaging their refined parameters, improves the final classification performance of the ensemble model.

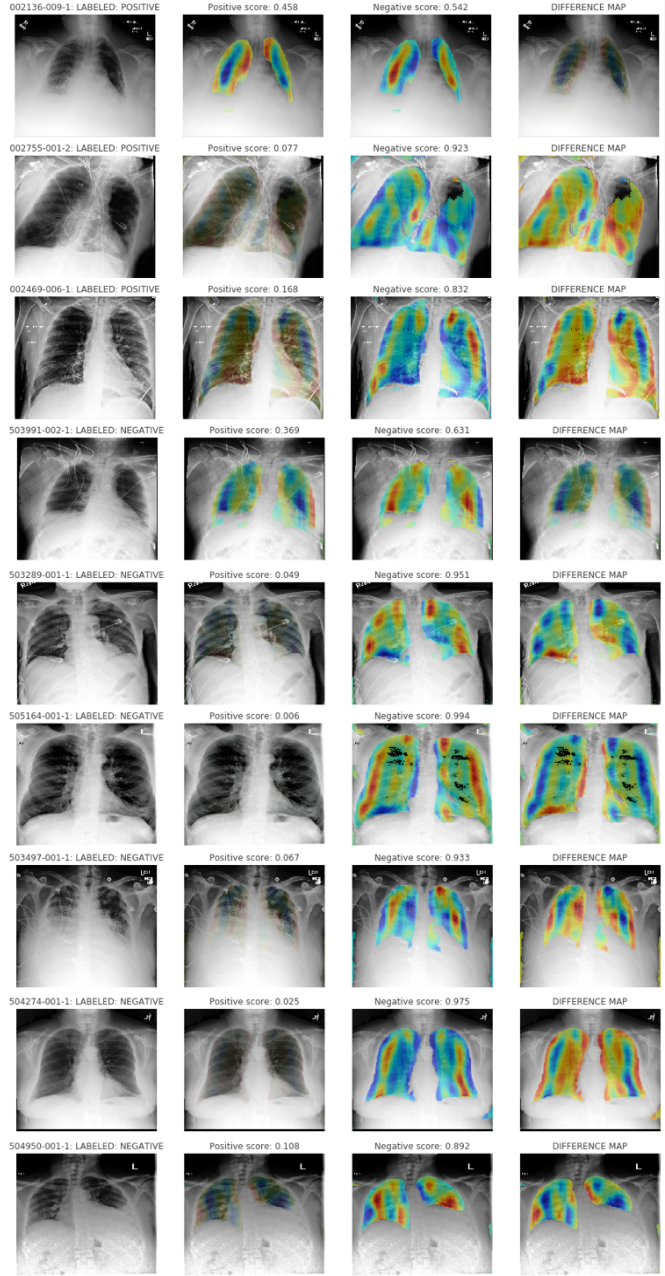

**Fig. 13.** Ensemble Module 2 predictions and Grad-CAM maps for some Covid + (top 3 rows) and Covid – (bottom 6 rows) CXRs in the HFHS test set. Panels of each row, from left to right are as in Fig. 12. The 1st row shows an uncertain assignment by the module as Covid +: consequently the difference Grad-CAM map is almost featureless. The 2nd and 3rd rows show CXRs labeled positive by the radiologist, but negative by Module 2. The bottom 6 rows shows CXRs labeled negative by the radiologist, and similarly assigned by Module 2, but with different degrees of certainty.

One important factor to consider in the evaluation of the performance of CXR-Net with the HFHS database of Covid + and Covid – CXRs, is that the Covid – subset was selected out of the large archive of CXRs maintained at HFHS only because the images were from pre-Covid era, regardless of whether they were normal or pathological. For this reason, the Covid – subset used in this study contains a large number of images of lungs affected by pathologies other than SARS-CoV-2. Thus, during its training Module 2 has learned to differentiate between Covid and anything that is non-Covid (normal or other forms of pneumonia). While this feature of the current CXR-Net trained model allows for a fast identification of Covid + cases based on a simple binary choice, it is expected to increase the number of *false negative* attributions, leading to a lower *recall* (**Table 2**) in comparison to other AI systems that have been developed for the same purpose (i.e., Covid-Net [4], DeepCOVID-XR [5]). However, the pipeline of CXR-Net can be easily modified by increasing the number of channels of the last convolutional layer of Module 2 leading into the Softmax activation, in order to allow a more fine-grained classification of CXRs into normal and various types of lung pathologies. The basic architecture of both modules of CXR-Net is derived from that of Res-CR-Net, a type of NN that was shown to be particularly effective in the multi-class semantic segmentation of microscopy images [6]. Thus, we expect CXR-Net to be equally effective in a multi-class identification of lung pathologies from A/P CXRs.

In its current form CXR-Net does not make use of additional information about the patients age, sex, chemistry/hematology data, and medical history. This additional information can be easily added a 3$^{rd}$ input scalar vector to Module 2, processed by a small Fully Connected NN, and ultimately merged into the Softmax layer. Module 1 is currently pre-trained on public databases, and thus is simply acting as an image pre-processing step for CXRs. Future versions of CXR-Net will simplify the current 2-steps pipeline by including Module 1 in the ensemble model as a pretrained layer with fixed parameters.

## Authors Contributions

[a]HA, SL, LP, and DLG developed the architecture of CXR-Net. [b]HA and BH, downloaded the JMS, MC, SH, and V7-Darwin databases of CXRs and corresponding lung masks from their public repositories, and prepared the local JMS and V7 datasets for use by CXR-Net. DLG and HSZ designed the study and wrote/edited the manuscript. All authors participated in discussions and proofreading of the manuscript.

[a]Hassan Abdallah, [b]Haikal Abdulah

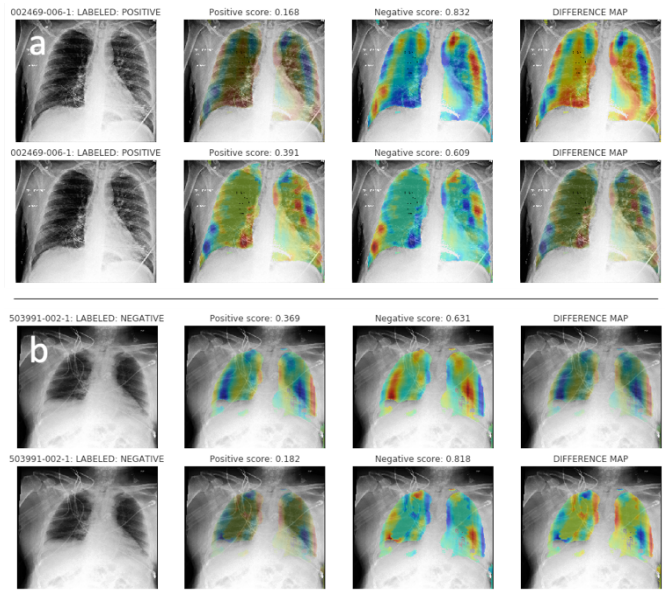

**Fig. 14**. Ensemble Module 2 predictions and Grad-CAM maps for some of the CXRs shown in Figs. 12,13, when CXR regions below a given threshold value are not included in the Global Average Pooling of the final convolutional layer. In each panels the top row is the saliency map obtained without thresholding, and the bottom row is the map obtained with thresholding. *Panel a*: an example of Covid + CXR in which the positive scoring is increased by thresholding. *Panel b*: an example of Covid – CXR in which the negative score is increased by thresholding.

## Software

Source code for CXR-Net is deposited at https://github.com/dgattiwsu/CXR-Net.

## ORCID

Domenico Gatti: 0000-0002-6357-3530
Hassan Abdallah: 0000-0003-1256-0446

## Acknowledgements

This study was supported by the WSU President's Research Enhancement Program in Computational Biology (DLG).